\documentclass[twocolumn,a4paper,superscriptaddress,reprint,aps,prb]{revtex4-1}

\usepackage{amsmath}
\usepackage{graphicx}
\usepackage[english]{babel}
\usepackage{calc}

\begin{document}

\title{Interplay between ferroic orders at the FeRh/BaTiO$_3$ interfaces}

\author{Viktoria V. Ivanovskaya}
\email{v.ivanovskaya@gmail.com}
\affiliation{Unit\'{e} Mixte de Physique CNRS-Thales, 1 Avenue Augustin
Fresnel,
91767 Palaiseau,
France and Universit\'{e} Paris-Sud, 91405 Orsay, France}

\author{Alberto Zobelli}
\affiliation{Laboratoire de Physique des Solides, Universit\'{e} Paris-Sud,
CNRS UMR 8502, F-91405,
Orsay, France}

\author{Alexandre Gloter}
\affiliation{Laboratoire de Physique des Solides, Universit\'{e} Paris-Sud,
CNRS UMR 8502, F-91405, Orsay, France}

\author{Manuel Bibes}
\affiliation{Unit\'{e} Mixte de Physique CNRS-Thales, 1 Avenue Augustin
Fresnel, 91767 Palaiseau, France and Universit\'{e} Paris-Sud, 91405 Orsay,
France}

\author{Agn\`es Barth\'el\'emy}
\affiliation{Unit\'{e} Mixte de Physique CNRS-Thales, 1 Avenue Augustin
Fresnel, 91767 Palaiseau, France and Universit\'{e} Paris-Sud, 91405 Orsay,
France}

\begin{abstract}

It has been recently demonstrated that the magnetic state of FeRh can be 
controlled by electric fields in FeRh/BaTiO$_{\text{3}}$ heterostructures [R.O. 
Cherifi \emph {et al.} Nature Mater. \textbf{13}, 345 (2014)]. 
Voltage-controlled changes in the ferroelastic domain structure of BaTiO$_3$ 
appeared to drive this effect, with charge accumulation and depletion due to 
ferroelectricity playing a more elusive role. To make this electric-field 
control of magnetic order non-volatile, the contribution of ferroelectric 
field-effect must be further enhanced, which requires understanding the 
details of the interface between FeRh and BaTiO$_3$. Here we report on the 
atomic structure and electron screening at this interface through density 
functional theory simulations. We relate different screening capabilities for 
the antiferromagnetic and ferromagnetic states of FeRh to different density of 
states at the Fermi level of corresponding bulk structures. We predict that the 
stability of the ferroelectric state in adjacent very thin BaTiO$_3$ films will 
be affected by magnetic order in FeRh. This control of ferroelectricity by 
magnetism can be viewed as the reciprocal effect of the voltage-controlled 
magnetic order previously found for this system.

\end{abstract}
\maketitle

\section{Introduction}\label{Intro}

The equilibrium ordered \textit{bcc} B2 FeRh alloy exhibits a first-order 
antiferromagnetic (AFM) to ferromagnetic (FM) phase transition slightly above 
room temperature.\cite{Fallot, muldawer} Recent achievements in the preparation 
of ordered alloy thin films combined with their potential technological 
applications (heat assisted magnetic recording or microelectromechanical 
devices, etc.) have stimulated a renewed interest for this 
material.\cite{fan2010ferromagnetism, bordel2012fe, bordel2, 
0022-3727-46-16-162002, de2013hall, gray2012electronic, miyanaga2009local, 
Inoue, Mancini, Baldasseroni, Vescovo} Experimentally it has been shown that in 
thin films the AFM to FM transition is highly sensitive to the film thickness 
and composition, heat treatment, magnetic field, pressure, etc.\cite{Han, 
kim2009surface, Han2, van1999compositional, 
cao2008magnetization,suzuki2009stability,Hillion,thiele2004magnetic, Staunton}

The capability to electrically switch the magnetic state of FeRh at a limited 
energy cost makes this material of the highest interest for potential spintronic 
applications. Very recently  it has been demonstrated that in 
FeRh/BaTiO$_{\text{3}}$ heterostructures a moderate electric field changing the 
BaTiO$_{\text{3}}$ ferroelastic state can produce a giant magnetization 
variation resulting from an AFM to a FM first order transition in the FeRh 
slab.\cite{Cherify-14} The effect occurs just above room temperature and it is 
mostly driven by voltage-induced strain from the BaTiO$_{\text{3}}$ substrate 
transfered to the FeRh. This effect appears to affect the FeRh film, but it has 
been suggested that an additional contribution to the magnetic phase transition 
may arise from more local interface effects due to charge accumulation and 
depletion.\cite{Cherify-14}

To make the voltage-induced change in the magnetic order non-volatile, such 
electronic effects must be made dominant over strain effects, which requires a 
detailed understanding of the subtle interplay between magnetism and the local 
electronic structure at the FeRh/BaTiO$_3$ interface. This is also especially 
important for future tunneling devices based on ferroelectric barriers and 
metamagnetic compounds such as FeRh as electrode. A common difficulty 
encountered while working with ferroelectric tunnel junctions is the 
preservation of ferroelectricity at the very low thickness of a tunnel 
barrier.\cite{juncekranature, garcia2014ferroelectric}  Such constrain was 
not present in Ref.\cite{Cherify-14} where FeRh films were grown onto 
BaTiO$_3$ substrate. An accurate description of the atomic structure of 
multiferroic thin film interfaces combined to an analysis of local electronic 
structure and screening at the metal electrodes is thus required to optimize 
their design.

In the present work we report the local electronic structure and electron 
screening at FeRh/BaTiO$_3$ interfaces obtained through density functional 
theory simulations. We show that different density of states at the Fermi level 
of bulk structures yield different screening capabilities for the AFM and FM 
phases. 
This effect is then related to the the critical thickness of the metal 
electrodes in order to stabilize the ferroelectric polarization.

\section{Computational details}

Spin-polarized density functional calculations were performed under the local 
density approximation as implemented in the AIMPRO code.\cite{Aimp-2, Aimpro} 
Relativistic pseudo-potentials were generated using the 
Hartwingster-Goedecker-Hutter scheme.\cite{Hgh-98} As basis sets, 50 independent 
Gaussian functions were used for iron and rhodium, 40 for titanium and oxygen 
and 20 for barium atoms. All atoms in a supercell were optimized using a 
conjugate gradient scheme, starting from the ferromagnetic or, alternatively, 
from the antiferromagnetic configuration in the FeRh slab, until the forces 
become less than 10$^{-4}$eV/{\AA}. Electronic structure convergence was ensured 
for each supercell by  using a $9\times 9 \times 1$ k points mesh generated from 
a Monkhorst-Pack set sampling of the Brillouin zone\cite{Monkhorst-Pack} and 
constraining the energy difference in the self-consistent cycle to be below 
10$^{-7}$ Hartree. 

\section{Results and discussion}

\begin{figure}[bt]
\includegraphics[width=\columnwidth]{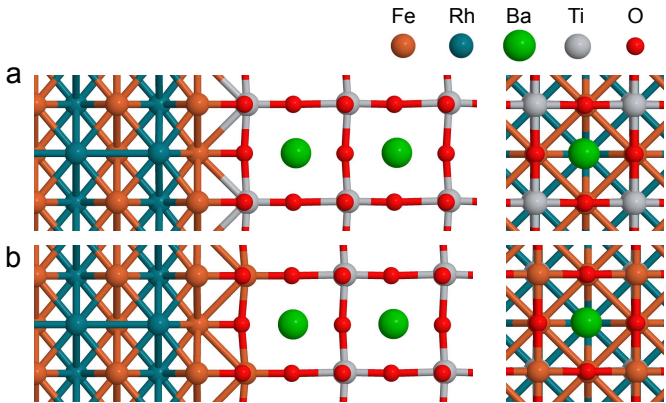}
\caption{Atomic model of the FeRh/BaTiO$_{\text{3}}$ interface with
(\textbf{a}) Fe interface plane and (\textbf{b}) FeO$_{\text{2}}$ interface
plane.}
\label{model}
\end{figure}

Atomic models have been built in a supercell approach considering 4.5 or 6.5 
unit cells for the metal slabs and 5.5, 7.5 or 11.5 unit cells for the 
perovskite slab, following the epitaxial relationship
(001)FeRh$\parallel$(001)BaTiO$_3$:[100]FeRh$\parallel$[110]BaTiO$_3$.
 We considered FeRh slabs terminated by both Fe-- or 
FeO$_{\text{2}}$-- planes at the interface with BaTiO$_{\text{3}}$ (note, that 
an FeO$_2$-type termination was found for the Fe/BaTiO$_3$ interface 
\cite{bocher2011atomic}). In the Fe-- terminated structure, metal atoms are 
located on top of the oxygen atoms of the perovskite surface (Fig. 
\ref{model}.a). From this configuration, the FeO$_{\text{2}}$-- terminated 
structure can be obtained by substituting in the --TiO$_{\text{2}}$ interfacial 
plane Ti by Fe leading to a further iron enrichment (Fig. \ref{model}.b). The 
two interfaces within the supercell have the same composition but different 
orientations with respect to the ferroelectric polarization, towards 
(P$_\text{up}$) or away from the interface (P$_\text{down}$). In our simulations 
the FeRh in-plane cell parameter \footnote{We find a cell parameters of 2.92 
{\AA} and 2.94 {\AA} for AFM FeRh and FM FeRh, respectively, in good agreement 
with experimental values\cite{Fallot, muldawer}} was constrained to that of 
BaTiO$_{\text{3}}$ (in our model 3.95{\AA}, obtained from bulk cell 
optimization) while the supercell out-of-plane parameter was free to 
relax. For the bulk FeRh structure, imposing epitaxy on the BaTiO$_{\text{3}}$ 
substrate, one leads to a tetragonal 
phase with the atomic and magnetic ordering of the FeRh alloy preserved, and a 
\textit{c/a} ratio of 1.15 for the AFM phase and 1.16 for the FM phase. These 
values differ from those of the bulk \emph{bcc} B2 phase but are very close to 
the experimental \text{c/a} ratio of the L1$_0$ phase.\cite{Takahashi, Oshima}

Similarly to what was observed for the cubic phase under hydrostatic 
compression, \cite{PhysRevB.67.064415, sandratskii2011magnetic} lattice 
distortions due to strain lead to only a slight decrease of the magnetization at 
the cubic to tetragonal phase transition. In the \textit{bcc} B2 structure, 
magnetic moments on the Fe and Rh atoms are respectively 3.21 and 0.00 $\mu$B in 
the AFM phase and 3.29 and 0.94 $\mu$B in FM phase, consistent with previous 
first-principles simulations.\cite{sandratskii2011magnetic, PhysRevB.67.064415, 
gu2005dominance} In the tetragonal structure promoted by the BaTiO$_{\text{3}}$ 
substrate we obtain for Fe and Rh atoms 3.07 $\mu$B and 0.0 $\mu$B in the AFM 
phase and 3.16 $\mu$B and 0.71 $\mu$B in the FM phase. This trend is consistent 
with experimental findings on somewhat lower magnetization in the tetragonal 
FeRh phase as compared to the cubic one.\cite{Miyajima}

We first discuss the stability of ferroelectricity in BaTiO$_3$ in the 
FeRh/BaTiO$_{\text{3}}$ system with Fe-- plane termination as a function of 
metal electrode thickness. It is well known that in ultrathin ferroelectric 
films strong depolarizing fields tend to suppress ferroelectricity and drive 
the system into a paraelectric state.  For instance, it has been calculated that 
the critical thickness for BaTiO$_{\text{3}}$ is 7 unit cells when interfaced 
with SrRuO$_3$.\cite{juncekranature} However, this result can not be 
generalized to different heterostructures since critical thicknesses are 
affected by the specific screening lengths of the chosen metal electrodes. 
Furthermore, the thickness of the metal is seldom discussed as an additional 
parameter influencing the ferroelectric stability, while it can have a strong 
influence for very thin electrodes.

\begin{figure}[bt]
\includegraphics[width=\columnwidth]{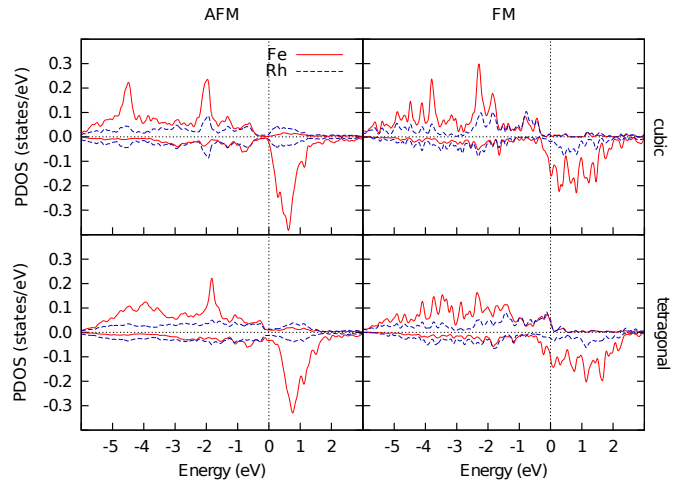}
\caption{Spin resolved Fe 3d and Rh 4d partial densities of states for AFM and
FM FeRh in cubic (top) and tetragonal (bottom) phases. Fermi level is set at
zero.}
\label{fig:PDOS}
\end{figure}

For FeRh electrodes, FM or AFM states should be considered as separated cases 
due to their different electronic properties that can be further modified by 
the tetragonal distortions imposed by BaTiO$_3$. Looking at the spin resolved 
partial densities of states (PDOS) projected at Fe and Rh sites (Fig. 
\ref{fig:PDOS}), the AFM FeRh in the cubic phase has a low density of states at 
the Fermi level which is only slightly increased when the system becomes 
tetragonal. The cubic FM FeRh has a high density of minority spins at the Fermi 
level with a spin polarization that is reduced in the tetragonal 
structure. The overall lower density of states at the Fermi level for AFM FeRh 
compared to FM FeRh is preserved both in the cubic and tetragonal phases. In 
both crystal structures, the density of states at the Fermi level is thus very 
different for the AFM and FM phases, consistent with experimental 
findings.\cite{de2013hall, gray2012electronic} FeRh thus has a different 
capability for charge accumulation or depletion depending on its magnetic state. 
As a consequence, higher screening lengths and thus higher depolarization fields 
are expected at AFM FeRh interfaces in contrast to FM FeRh interfaces.

\begin{figure}[tb]
\includegraphics[width=\columnwidth]{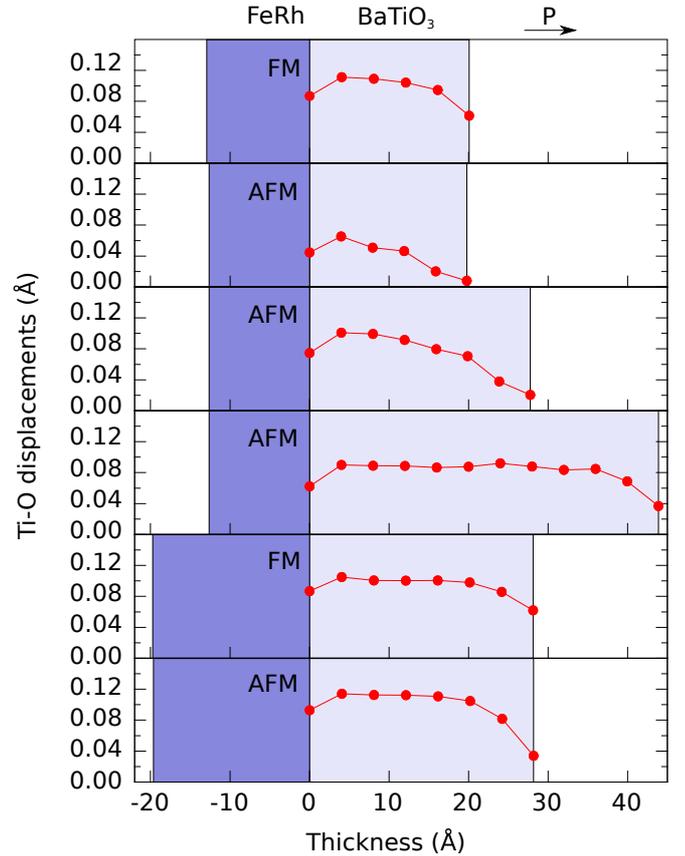}
\caption{Displacements of Ti atoms related to the O atoms at TiO$_2$--planes for
heterostructures with different FM and AFM FeRh and BaTiO$_3$ thicknesses (red 
curves). The thicknesses of 
ferrolectric and metallic slabs are marked by rectangles. The arrow shows the 
direction of the polarization in the ferroelectric slab.}
\label{fig:disps}
\end{figure}

In Fig. \ref{fig:disps} we present the amplitude of the Ti-O displacements at 
TiO$_{\text{2}}$ planes (whose values are related to the ferroelectric 
polarization) as a function of the BaTiO$_{\text{3}}$ and FeRh slabs thicknesses 
for FM and AFM magnetic states. In a previous study, the critical thickness for 
BaTiO$_{\text{3}}$ slabs has been evaluated in about 18 {\AA} (4.5 unit cells) 
for  13 {\AA} thick Fe electrodes (9 Fe planes).\cite{duan2006predicted} Using 
similar ferroelectric and metal slabs thicknesses for the FeRh case, we find 
that Ti-O displacements in the middle section of the BaTiO$_{\text{3}}$ slab 
approach the bulk value (0.12 \AA) only when the FeRh electrode is in the FM  
magnetic state. A low rumpling and a loss of polarization (corresponding to 
about the half of the bulk displacements) is instead associated to the FeRh AFM 
magnetic state. This effect occurs even for thicker  BaTiO$_{\text{3}}$ slabs 
(30 {\AA} and 40 {\AA}) for which depolarizing fields are lower. The critical 
thickness of BaTiO$_{\text{3}}$ to preserve the ferroelectricity should occur 
above 40 {\AA} for AFM FeRh electrodes while it is less then 20 {\AA} for FM 
FeRh electrodes. In other words, the ferroelectric polarization of a BaTiO$_3$ 
film with a thickness between 20 and 40 {\AA} can be controlled by switching the 
magnetic state of the adjacent thin FeRh slab. This behavior can be viewed as a 
reciprocal effect of the voltage controlled magnetic order observed for such 
heterostructures.\cite{Cherify-14} Finally, for all structures considered we 
observe a higher rumpling of the polarization at the P$_\text{up}$ interface 
compared to the P$_\text{down}$ interface.

\begin{figure*}
\includegraphics[width=\textwidth]{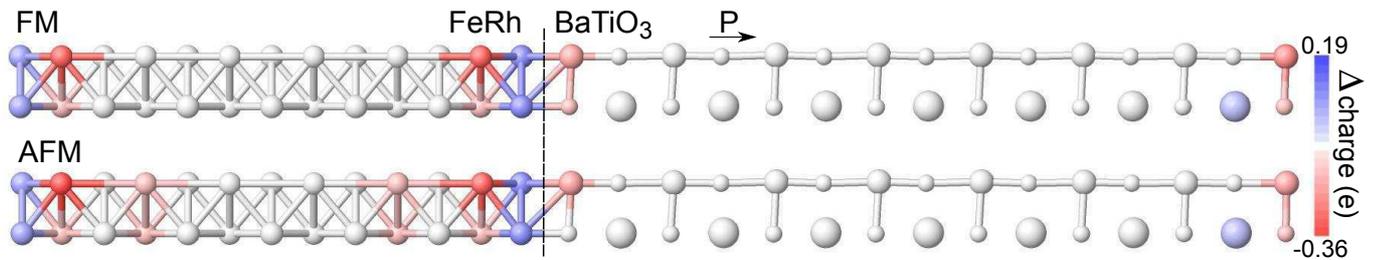}
\caption{Differential charge distribution for the FM (top) and AFM (bottom)
FeRh/BaTiO$_3$ heterostructures. Charges are calculated as the difference 
between
the induced charges at the interfaces and their bulk
counterparts; the polarization direction is indicated by the arrow.}
\label{charges}
\end{figure*}

The low Ti-O displacements observed for AFM FeRh interfaces are a consequence of 
the strong residual depolarizing fields due to incomplete charge screening. We 
evaluate charges for bulk and interface structures through a Mulliken 
population analysis (Tab. \ref{tab-charges} and Fig. \ref{charges}). In bulk 
FeRh we obtain a charge transfer of +0.71 $e$ from Rh to Fe for the AFM and 
+0.69 $e$ for the FM magnetic configurations (Tab. \ref{tab-charges}). 
Considering the BaTiO$_{\text{3}}$ slab in the paraelectric state (mirror 
symmetry imposed), iron is further positively charged at the interface due to 
its partial oxidation. For the BaTiO$_{\text{3}}$ ferroelectric state, the 
polarization screening variates the charge state at the 
P$_\text{up}$ and P$_\text{down}$ interfaces. For the P$_\text{up}$ interface a 
higher electron accumulation at the first Fe plane occurs for FM FeRh compared 
to AFM FeRh (Tab. \ref{tab-charges}).

Besides the first interfacial Fe plane, in Fig.\ref{charges} we represent the 
difference of charge with respect to bulk charge states for the whole 
heterostructures. In the case of FM FeRh the carrier density is high enough to 
provide a screening length of the order of only a few surface metal layers (Fig. 
\ref{charges}). For AFM FeRh, which is a worst metal as we discussed above, this 
length increases significantly and for thin metal layers it becomes of the same 
order of magnitude as the electrode thickness. A way to reduce the critical 
thickness for the ferroelectric can be achieved through a better screening by 
increasing the metal thickness. This explains the different critical thicknesses 
observed for the two magnetic phases (Figure \ref{fig:disps}).

\begin{table}[b]
\caption{Fe charge in bulk FeRh and charge difference from the bulk charge
for Fe atoms at the first atomic plane of the FeRh/BaTiO$_{\text{3}}$ 
interface.}
\label{tab-charges}
 \begin{tabular*}{\linewidth}{@{\extracolsep{\fill}}lcc}
  \hline
  \hline
   \multicolumn{3}{c}{Fe charge reference ($e$/atom) } \\
     & FM & AFM \\
     \hline
Bulk FeRh & 0.71 & 0.69 \\
   \hline
  \hline
 \multicolumn{3}{c}{Fe $\Delta$charge at the interface plane ($e$/atom)}\\
    & FM & AFM \\
\hline
 
 \hline
  BaTiO$_{\text{3}}$, paraelectric & 0.16 & 0.16 \\
  BaTiO$_{\text{3}}$, P$_\text{up}$ interface & 0.11 & 0.13 \\
  BaTiO$_{\text{3}}$, P$_\text{down}$ interface  & 0.17 & 0.17 \\
   \hline

  \hline
  \hline
 \end{tabular*}
\end{table}

We should consider also that the ferroelectric polarization in the presence of 
AFM FeRh slabs promotes a local transition to the FM phase solely of the first 
interfacial unit cell. Whereas bulk FM FeRh is 93 meV higher in energy than AFM 
FeRh, we find that this local magnetic switch at the P$_\text{up}$ interface 
lead to an energy gain of 100 meV per interface unit cell. Nevertheless, this 
more complex magnetic state does not affect trends in ferroelectric 
polarizations discussed above. However, the application of an electric field or 
the use of a ferroelectric with a higher polarization should promote this 
interfacial AFM to FM transition in a bigger volume close to the interface. This 
can have important consequences for spin-dependent transport in multiferroic 
tunnel junctions based on FeRh.

Finally, we comment on screening effects in the presence of a FeRh oxidized 
interface i.e. with a FeO$_2$--  interfacial plane. 
Charges in this last plane contribute also to the screening. In the above 
considered unoxidized case, there is a charge accumulation of  -0.71 $e$  for 
P$_\text{down}$ and -0.82  $e$ for P$_\text{up}$ at the terminal 
TiO$_{\text{2}}$-- plane (Fig. 
\ref{charges}). In the oxidized case, the FeO$_{\text{2}}$-- plane 
has a ferromagnetic order and the charge state is -0.91 $e$ at the 
P$_\text{down}$ interface and -1.05 $e$ at the P$_\text{up}$ interface, which 
results in an enhanced screening. Thus, charge accumulation at the successive 
FeRh planes is reduced and in this case the system in the AFM state is not 
further stabilized by switching the first unit cell to the FM state.

\section{Conclusions}

In this paper, via density functional theory simulations we have provided 
insights on the role of the electrode magnetic state in stabilizing the 
ferroelectric polarization for the FeRh/BaTiO$_3$ system.
In particular, we have shown that screening is higher 
for FM FeRh than AFM FeRh. This behavior is in agreement with the differences on 
the near Fermi energy density of states of the two magnetic phases observed in 
both cubic and tetragonal bulk lattices: AFM FeRh can be described as a  
worst metal than FM FeRh. For thin FeRh/BaTiO$_3$ heterostructures the 
thicknesses of both the ferroelectric and ferromagnetic layers contribute to the 
stability of the ferroelectric phase. Indeed at a given ferroelectric thickness, 
associated with a specific depolarizing field, it is possible to define a 
critical thickness of the ferromagnetic electrode below which the screening is 
incomplete. The different screening capabilities of the FM FeRh and AFM FeRh 
phases lead then to different critical thicknesses.

In a previous work \cite{Cherify-14} we presented experimental evidences for a 
control of the FeRh magnetic state in the FeRh/BaTiO$_{\text{3}}$ 
heterostructure through the BaTiO$_{\text{3}}$ ferroelectric polarization. The 
theoretical results presented here suggest a mechanism where the switch between 
magnetic states of a thin FeRh electrode would lead to complete or uncomplete 
screening and thus to changes in the ferroelectric polarization (and possibly 
Curie temperature) of the BaTiO$_{\text{3}}$. This mechanism should be general, 
which suggests that changes in the ferroelectric properties should occur in 
heterostructures combining a ferroelectric with a material hosting phases with 
different densities of states. Such systems are ubiquitous in the perovskite 
family (for instance manganites or nickelates with sharp metal-insulator 
transitions\cite{catalan2008progress, medarde1997structural, 
RevModPhys.70.1039}) that could be 
easily combined with BaTiO$_3$ or 
other ferroelectrics.

\acknowledgements

This work received financial support from the French Agence Nationale de la 
Recherche through project NOMILOPS (ANR-11-BS10-0016) and the European Research 
Council Advanced Grant FEMMES (contract no. 267579). We would like to thank 
L.C. Philips for usefull comments.


\begin{thebibliography}{41}%
\makeatletter
\providecommand \@ifxundefined [1]{%
 \@ifx{#1\undefined}
}%
\providecommand \@ifnum [1]{%
 \ifnum #1\expandafter \@firstoftwo
 \else \expandafter \@secondoftwo
 \fi
}%
\providecommand \@ifx [1]{%
 \ifx #1\expandafter \@firstoftwo
 \else \expandafter \@secondoftwo
 \fi
}%
\providecommand \natexlab [1]{#1}%
\providecommand \enquote  [1]{``#1''}%
\providecommand \bibnamefont  [1]{#1}%
\providecommand \bibfnamefont [1]{#1}%
\providecommand \citenamefont [1]{#1}%
\providecommand \href@noop [0]{\@secondoftwo}%
\providecommand \href [0]{\begingroup \@sanitize@url \@href}%
\providecommand \@href[1]{\@@startlink{#1}\@@href}%
\providecommand \@@href[1]{\endgroup#1\@@endlink}%
\providecommand \@sanitize@url [0]{\catcode `\\12\catcode `\$12\catcode
  `\&12\catcode `\#12\catcode `\^12\catcode `\_12\catcode `\%12\relax}%
\providecommand \@@startlink[1]{}%
\providecommand \@@endlink[0]{}%
\providecommand \url  [0]{\begingroup\@sanitize@url \@url }%
\providecommand \@url [1]{\endgroup\@href {#1}{\urlprefix }}%
\providecommand \urlprefix  [0]{URL }%
\providecommand \Eprint [0]{\href }%
\providecommand \doibase [0]{http://dx.doi.org/}%
\providecommand \selectlanguage [0]{\@gobble}%
\providecommand \bibinfo  [0]{\@secondoftwo}%
\providecommand \bibfield  [0]{\@secondoftwo}%
\providecommand \translation [1]{[#1]}%
\providecommand \BibitemOpen [0]{}%
\providecommand \bibitemStop [0]{}%
\providecommand \bibitemNoStop [0]{.\EOS\space}%
\providecommand \EOS [0]{\spacefactor3000\relax}%
\providecommand \BibitemShut  [1]{\csname bibitem#1\endcsname}%
\let\auto@bib@innerbib\@empty
\bibitem [{\citenamefont {Fallot}(1938)}]{Fallot}%
  \BibitemOpen
  \bibfield  {author} {\bibinfo {author} {\bibfnamefont {M.}~\bibnamefont
  {Fallot}},\ }\href@noop {} {\bibfield  {journal} {\bibinfo  {journal} {Ann.
  Phys.}\ }\textbf {\bibinfo {volume} {10}},\ \bibinfo {pages} {291} (\bibinfo
  {year} {1938})}\BibitemShut {NoStop}%
\bibitem [{\citenamefont {Muldawer}\ and\ \citenamefont
  {de~Bergevin}(1961)}]{muldawer}%
  \BibitemOpen
  \bibfield  {author} {\bibinfo {author} {\bibfnamefont {L.}~\bibnamefont
  {Muldawer}}\ and\ \bibinfo {author} {\bibfnamefont {F.}~\bibnamefont
  {de~Bergevin}},\ }\href@noop {} {\bibfield  {journal} {\bibinfo  {journal}
  {J. Chem. Phys.}\ }\textbf {\bibinfo {volume} {35}},\ \bibinfo {pages} {1904}
  (\bibinfo {year} {1961})}\BibitemShut {NoStop}%
\bibitem [{\citenamefont {Fan}\ \emph {et~al.}(2010)\citenamefont {Fan},
  \citenamefont {Kinane}, \citenamefont {Charlton}, \citenamefont {Dorner},
  \citenamefont {Ali}, \citenamefont {de~Vries}, \citenamefont {Brydson},
  \citenamefont {Marrows}, \citenamefont {Hickey}, \citenamefont {Arena} \emph
  {et~al.}}]{fan2010ferromagnetism}%
  \BibitemOpen
  \bibfield  {author} {\bibinfo {author} {\bibfnamefont {R.}~\bibnamefont
  {Fan}}, \bibinfo {author} {\bibfnamefont {C.}~\bibnamefont {Kinane}},
  \bibinfo {author} {\bibfnamefont {T.}~\bibnamefont {Charlton}}, \bibinfo
  {author} {\bibfnamefont {R.}~\bibnamefont {Dorner}}, \bibinfo {author}
  {\bibfnamefont {M.}~\bibnamefont {Ali}}, \bibinfo {author} {\bibfnamefont
  {M.}~\bibnamefont {de~Vries}}, \bibinfo {author} {\bibfnamefont
  {R.}~\bibnamefont {Brydson}}, \bibinfo {author} {\bibfnamefont
  {C.}~\bibnamefont {Marrows}}, \bibinfo {author} {\bibfnamefont
  {B.}~\bibnamefont {Hickey}}, \bibinfo {author} {\bibfnamefont
  {D.}~\bibnamefont {Arena}},  \emph {et~al.},\ }\href@noop {} {\bibfield
  {journal} {\bibinfo  {journal} {Phys. Rev. B}\ }\textbf {\bibinfo {volume}
  {82}},\ \bibinfo {pages} {184418} (\bibinfo {year} {2010})}\BibitemShut
  {NoStop}%
\bibitem [{\citenamefont {Bordel}\ \emph {et~al.}(2012)\citenamefont {Bordel},
  \citenamefont {Juraszek}, \citenamefont {Cooke}, \citenamefont
  {Baldasseroni}, \citenamefont {Mankovsky}, \citenamefont {Min{\'a}r},
  \citenamefont {Ebert}, \citenamefont {Moyerman}, \citenamefont {Fullerton},\
  and\ \citenamefont {Hellman}}]{bordel2012fe}%
  \BibitemOpen
  \bibfield  {author} {\bibinfo {author} {\bibfnamefont {C.}~\bibnamefont
  {Bordel}}, \bibinfo {author} {\bibfnamefont {J.}~\bibnamefont {Juraszek}},
  \bibinfo {author} {\bibfnamefont {D.~W.}\ \bibnamefont {Cooke}}, \bibinfo
  {author} {\bibfnamefont {C.}~\bibnamefont {Baldasseroni}}, \bibinfo {author}
  {\bibfnamefont {S.}~\bibnamefont {Mankovsky}}, \bibinfo {author}
  {\bibfnamefont {J.}~\bibnamefont {Min{\'a}r}}, \bibinfo {author}
  {\bibfnamefont {H.}~\bibnamefont {Ebert}}, \bibinfo {author} {\bibfnamefont
  {S.}~\bibnamefont {Moyerman}}, \bibinfo {author} {\bibfnamefont
  {E.}~\bibnamefont {Fullerton}}, \ and\ \bibinfo {author} {\bibfnamefont
  {F.}~\bibnamefont {Hellman}},\ }\href@noop {} {\bibfield  {journal} {\bibinfo
   {journal} {Phys. Rev. Lett.}\ }\textbf {\bibinfo {volume} {109}},\ \bibinfo
  {pages} {117201} (\bibinfo {year} {2012})}\BibitemShut {NoStop}%
\bibitem [{\citenamefont {Cooke}\ \emph {et~al.}(2012)\citenamefont {Cooke},
  \citenamefont {Hellman}, \citenamefont {Baldasseroni}, \citenamefont
  {Bordel}, \citenamefont {Moyerman},\ and\ \citenamefont
  {Fullerton}}]{bordel2}%
  \BibitemOpen
  \bibfield  {author} {\bibinfo {author} {\bibfnamefont {D.~W.}\ \bibnamefont
  {Cooke}}, \bibinfo {author} {\bibfnamefont {F.}~\bibnamefont {Hellman}},
  \bibinfo {author} {\bibfnamefont {C.}~\bibnamefont {Baldasseroni}}, \bibinfo
  {author} {\bibfnamefont {C.}~\bibnamefont {Bordel}}, \bibinfo {author}
  {\bibfnamefont {S.}~\bibnamefont {Moyerman}}, \ and\ \bibinfo {author}
  {\bibfnamefont {E.~E.}\ \bibnamefont {Fullerton}},\ }\href@noop {} {\bibfield
   {journal} {\bibinfo  {journal} {Phys. Rev. Lett.}\ }\textbf {\bibinfo
  {volume} {109}},\ \bibinfo {pages} {255901} (\bibinfo {year}
  {2012})}\BibitemShut {NoStop}%
\bibitem [{\citenamefont {Loving}\ \emph {et~al.}(2013)\citenamefont {Loving},
  \citenamefont {Jimenez-Villacorta}, \citenamefont {Kaeswurm}, \citenamefont
  {Arena}, \citenamefont {Marrows},\ and\ \citenamefont
  {Lewis}}]{0022-3727-46-16-162002}%
  \BibitemOpen
  \bibfield  {author} {\bibinfo {author} {\bibfnamefont {M.}~\bibnamefont
  {Loving}}, \bibinfo {author} {\bibfnamefont {F.}~\bibnamefont
  {Jimenez-Villacorta}}, \bibinfo {author} {\bibfnamefont {B.}~\bibnamefont
  {Kaeswurm}}, \bibinfo {author} {\bibfnamefont {D.~A.}\ \bibnamefont {Arena}},
  \bibinfo {author} {\bibfnamefont {C.~H.}\ \bibnamefont {Marrows}}, \ and\
  \bibinfo {author} {\bibfnamefont {L.~H.}\ \bibnamefont {Lewis}},\ }\href
  {http://stacks.iop.org/0022-3727/46/i=16/a=162002} {\bibfield  {journal}
  {\bibinfo  {journal} {J. Phys. D: Appl. Phys.}\ }\textbf {\bibinfo {volume}
  {46}},\ \bibinfo {pages} {162002} (\bibinfo {year} {2013})}\BibitemShut
  {NoStop}%
\bibitem [{\citenamefont {de~Vries}\ \emph {et~al.}(2013)\citenamefont
  {de~Vries}, \citenamefont {Loving}, \citenamefont {Mihai}, \citenamefont
  {Lewis}, \citenamefont {Heiman},\ and\ \citenamefont {Marrows}}]{de2013hall}%
  \BibitemOpen
  \bibfield  {author} {\bibinfo {author} {\bibfnamefont {M.}~\bibnamefont
  {de~Vries}}, \bibinfo {author} {\bibfnamefont {M.}~\bibnamefont {Loving}},
  \bibinfo {author} {\bibfnamefont {A.}~\bibnamefont {Mihai}}, \bibinfo
  {author} {\bibfnamefont {L.}~\bibnamefont {Lewis}}, \bibinfo {author}
  {\bibfnamefont {D.}~\bibnamefont {Heiman}}, \ and\ \bibinfo {author}
  {\bibfnamefont {C.}~\bibnamefont {Marrows}},\ }\href@noop {} {\bibfield
  {journal} {\bibinfo  {journal} {New J. Phys.}\ }\textbf {\bibinfo {volume}
  {15}},\ \bibinfo {pages} {013008} (\bibinfo {year} {2013})}\BibitemShut
  {NoStop}%
\bibitem [{\citenamefont {Gray}\ \emph {et~al.}(2012)\citenamefont {Gray},
  \citenamefont {Cooke}, \citenamefont {Kr{\"u}ger}, \citenamefont {Bordel},
  \citenamefont {Kaiser}, \citenamefont {Moyerman}, \citenamefont {Fullerton},
  \citenamefont {Ueda}, \citenamefont {Yamashita}, \citenamefont {Gloskovskii}
  \emph {et~al.}}]{gray2012electronic}%
  \BibitemOpen
  \bibfield  {author} {\bibinfo {author} {\bibfnamefont {A.~X.}\ \bibnamefont
  {Gray}}, \bibinfo {author} {\bibfnamefont {D.~W.}\ \bibnamefont {Cooke}},
  \bibinfo {author} {\bibfnamefont {P.}~\bibnamefont {Kr{\"u}ger}}, \bibinfo
  {author} {\bibfnamefont {C.}~\bibnamefont {Bordel}}, \bibinfo {author}
  {\bibfnamefont {A.~M.}\ \bibnamefont {Kaiser}}, \bibinfo {author}
  {\bibfnamefont {S.}~\bibnamefont {Moyerman}}, \bibinfo {author}
  {\bibfnamefont {E.~E.}\ \bibnamefont {Fullerton}}, \bibinfo {author}
  {\bibfnamefont {S.}~\bibnamefont {Ueda}}, \bibinfo {author} {\bibfnamefont
  {Y.}~\bibnamefont {Yamashita}}, \bibinfo {author} {\bibfnamefont
  {A.}~\bibnamefont {Gloskovskii}},  \emph {et~al.},\ }\href@noop {} {\bibfield
   {journal} {\bibinfo  {journal} {Phys. Rev. Lett.}\ }\textbf {\bibinfo
  {volume} {108}},\ \bibinfo {pages} {257208} (\bibinfo {year}
  {2012})}\BibitemShut {NoStop}%
\bibitem [{\citenamefont {Miyanaga}\ \emph {et~al.}(2009)\citenamefont
  {Miyanaga}, \citenamefont {Itoga}, \citenamefont {Okazaki},\ and\
  \citenamefont {Nitta}}]{miyanaga2009local}%
  \BibitemOpen
  \bibfield  {author} {\bibinfo {author} {\bibfnamefont {T.}~\bibnamefont
  {Miyanaga}}, \bibinfo {author} {\bibfnamefont {T.}~\bibnamefont {Itoga}},
  \bibinfo {author} {\bibfnamefont {T.}~\bibnamefont {Okazaki}}, \ and\
  \bibinfo {author} {\bibfnamefont {K.}~\bibnamefont {Nitta}},\ }\href@noop {}
  {\bibfield  {journal} {\bibinfo  {journal} {J. Phys.: Conf. Ser.}\ }\textbf
  {\bibinfo {volume} {190}},\ \bibinfo {pages} {012097} (\bibinfo {year}
  {2009})}\BibitemShut {NoStop}%
\bibitem [{\citenamefont {Inoue}\ \emph {et~al.}(2008)\citenamefont {Inoue},
  \citenamefont {Phuoc}, \citenamefont {Cao}, \citenamefont {Nam}, ,
  \citenamefont {Ko},\ and\ \citenamefont {Suzuki}}]{Inoue}%
  \BibitemOpen
  \bibfield  {author} {\bibinfo {author} {\bibfnamefont {S.}~\bibnamefont
  {Inoue}}, \bibinfo {author} {\bibfnamefont {N.~N.}\ \bibnamefont {Phuoc}},
  \bibinfo {author} {\bibfnamefont {J.}~\bibnamefont {Cao}}, \bibinfo {author}
  {\bibfnamefont {N.~T.}\ \bibnamefont {Nam}}, , \bibinfo {author}
  {\bibfnamefont {H.~Y.~Y.}\ \bibnamefont {Ko}}, \ and\ \bibinfo {author}
  {\bibfnamefont {T.}~\bibnamefont {Suzuki}},\ }\href@noop {} {\bibfield
  {journal} {\bibinfo  {journal} {J. Appl. Phys.}\ }\textbf {\bibinfo {volume}
  {103}},\ \bibinfo {pages} {07B312} (\bibinfo {year} {2008})}\BibitemShut
  {NoStop}%
\bibitem [{\citenamefont {Mancini}\ \emph {et~al.}(2013)\citenamefont
  {Mancini}, \citenamefont {Pressacco}, \citenamefont {Haertinger},
  \citenamefont {Fullerton}, \citenamefont {Suzuki}, \citenamefont
  {Woltersdorf},\ and\ \citenamefont {Back}}]{Mancini}%
  \BibitemOpen
  \bibfield  {author} {\bibinfo {author} {\bibfnamefont {E.}~\bibnamefont
  {Mancini}}, \bibinfo {author} {\bibfnamefont {F.}~\bibnamefont {Pressacco}},
  \bibinfo {author} {\bibfnamefont {M.}~\bibnamefont {Haertinger}}, \bibinfo
  {author} {\bibfnamefont {E.~E.}\ \bibnamefont {Fullerton}}, \bibinfo {author}
  {\bibfnamefont {T.}~\bibnamefont {Suzuki}}, \bibinfo {author} {\bibfnamefont
  {G.}~\bibnamefont {Woltersdorf}}, \ and\ \bibinfo {author} {\bibfnamefont
  {C.~H.}\ \bibnamefont {Back}},\ }\href@noop {} {\bibfield  {journal}
  {\bibinfo  {journal} {J.Phys. D: Appl. Phys.}\ }\textbf {\bibinfo {volume}
  {46}},\ \bibinfo {pages} {245302} (\bibinfo {year} {2013})}\BibitemShut
  {NoStop}%
\bibitem [{\citenamefont {Baldasseroni}\ \emph {et~al.}(2012)\citenamefont
  {Baldasseroni}, \citenamefont {Bordel}, \citenamefont {Gray}, \citenamefont
  {Kaiser}, \citenamefont {Kronast}, \citenamefont {Herrero-Albillos},
  \citenamefont {Schneider}, \citenamefont {Fadley},\ and\ \citenamefont
  {Hellman}}]{Baldasseroni}%
  \BibitemOpen
  \bibfield  {author} {\bibinfo {author} {\bibfnamefont {C.}~\bibnamefont
  {Baldasseroni}}, \bibinfo {author} {\bibfnamefont {C.}~\bibnamefont
  {Bordel}}, \bibinfo {author} {\bibfnamefont {A.}~\bibnamefont {Gray}},
  \bibinfo {author} {\bibfnamefont {A.}~\bibnamefont {Kaiser}}, \bibinfo
  {author} {\bibfnamefont {F.}~\bibnamefont {Kronast}}, \bibinfo {author}
  {\bibfnamefont {J.}~\bibnamefont {Herrero-Albillos}}, \bibinfo {author}
  {\bibfnamefont {C.}~\bibnamefont {Schneider}}, \bibinfo {author}
  {\bibfnamefont {C.}~\bibnamefont {Fadley}}, \ and\ \bibinfo {author}
  {\bibfnamefont {F.}~\bibnamefont {Hellman}},\ }\href@noop {} {\bibfield
  {journal} {\bibinfo  {journal} {Appl. Phys. Lett.}\ }\textbf {\bibinfo
  {volume} {100}},\ \bibinfo {pages} {262401} (\bibinfo {year}
  {2012})}\BibitemShut {NoStop}%
\bibitem [{\citenamefont {Lee}\ \emph {et~al.}(2010)\citenamefont {Lee},
  \citenamefont {Vescovo}, \citenamefont {Plucinski}, \citenamefont
  {Schneider},\ and\ \citenamefont {Kao}}]{Vescovo}%
  \BibitemOpen
  \bibfield  {author} {\bibinfo {author} {\bibfnamefont {J.-S.}\ \bibnamefont
  {Lee}}, \bibinfo {author} {\bibfnamefont {E.}~\bibnamefont {Vescovo}},
  \bibinfo {author} {\bibfnamefont {L.}~\bibnamefont {Plucinski}}, \bibinfo
  {author} {\bibfnamefont {C.~M.}\ \bibnamefont {Schneider}}, \ and\ \bibinfo
  {author} {\bibfnamefont {C.-C.}\ \bibnamefont {Kao}},\ }\href@noop {}
  {\bibfield  {journal} {\bibinfo  {journal} {Phys. Rev. B.}\ }\textbf
  {\bibinfo {volume} {82}},\ \bibinfo {pages} {224410} (\bibinfo {year}
  {2010})}\BibitemShut {NoStop}%
\bibitem [{\citenamefont {Han}\ \emph {et~al.}(2013{\natexlab{a}})\citenamefont
  {Han}, \citenamefont {Qiu}, \citenamefont {Yap}, \citenamefont {Luo},
  \citenamefont {Kanbe}, \citenamefont {Shige}, \citenamefont {Laughlin},\ and\
  \citenamefont {Zhu}}]{Han}%
  \BibitemOpen
  \bibfield  {author} {\bibinfo {author} {\bibfnamefont {G.~C.}\ \bibnamefont
  {Han}}, \bibinfo {author} {\bibfnamefont {J.~J.}\ \bibnamefont {Qiu}},
  \bibinfo {author} {\bibfnamefont {Q.}~\bibnamefont {Yap}}, \bibinfo {author}
  {\bibfnamefont {P.}~\bibnamefont {Luo}}, \bibinfo {author} {\bibfnamefont
  {T.}~\bibnamefont {Kanbe}}, \bibinfo {author} {\bibfnamefont
  {T.}~\bibnamefont {Shige}}, \bibinfo {author} {\bibfnamefont {D.~E.}\
  \bibnamefont {Laughlin}}, \ and\ \bibinfo {author} {\bibfnamefont {J.-G.}\
  \bibnamefont {Zhu}},\ }\href@noop {} {\bibfield  {journal} {\bibinfo
  {journal} {J. Appl. Phys.}\ }\textbf {\bibinfo {volume} {113}},\ \bibinfo
  {pages} {123909} (\bibinfo {year} {2013}{\natexlab{a}})}\BibitemShut
  {NoStop}%
\bibitem [{\citenamefont {Kim}\ \emph {et~al.}(2009)\citenamefont {Kim},
  \citenamefont {Ryan}, \citenamefont {Ding}, \citenamefont {Lewis},
  \citenamefont {Ali}, \citenamefont {Kinane}, \citenamefont {Hickey},
  \citenamefont {Marrows},\ and\ \citenamefont {Arena}}]{kim2009surface}%
  \BibitemOpen
  \bibfield  {author} {\bibinfo {author} {\bibfnamefont {J.}~\bibnamefont
  {Kim}}, \bibinfo {author} {\bibfnamefont {P.}~\bibnamefont {Ryan}}, \bibinfo
  {author} {\bibfnamefont {Y.}~\bibnamefont {Ding}}, \bibinfo {author}
  {\bibfnamefont {L.}~\bibnamefont {Lewis}}, \bibinfo {author} {\bibfnamefont
  {M.}~\bibnamefont {Ali}}, \bibinfo {author} {\bibfnamefont {C.}~\bibnamefont
  {Kinane}}, \bibinfo {author} {\bibfnamefont {B.}~\bibnamefont {Hickey}},
  \bibinfo {author} {\bibfnamefont {C.}~\bibnamefont {Marrows}}, \ and\
  \bibinfo {author} {\bibfnamefont {D.}~\bibnamefont {Arena}},\ }\href@noop {}
  {\bibfield  {journal} {\bibinfo  {journal} {Appl. Phys. Lett.}\ }\textbf
  {\bibinfo {volume} {95}},\ \bibinfo {pages} {222515} (\bibinfo {year}
  {2009})}\BibitemShut {NoStop}%
\bibitem [{\citenamefont {Han}\ \emph {et~al.}(2013{\natexlab{b}})\citenamefont
  {Han}, \citenamefont {Qiu}, \citenamefont {Yap}, \citenamefont {Luo},
  \citenamefont {Laughlin}, \citenamefont {Zhu}, \citenamefont {Kanbe},\ and\
  \citenamefont {Shige}}]{Han2}%
  \BibitemOpen
  \bibfield  {author} {\bibinfo {author} {\bibfnamefont {G.~C.}\ \bibnamefont
  {Han}}, \bibinfo {author} {\bibfnamefont {J.~J.}\ \bibnamefont {Qiu}},
  \bibinfo {author} {\bibfnamefont {Q.}~\bibnamefont {Yap}}, \bibinfo {author}
  {\bibfnamefont {P.}~\bibnamefont {Luo}}, \bibinfo {author} {\bibfnamefont
  {D.~E.}\ \bibnamefont {Laughlin}}, \bibinfo {author} {\bibfnamefont {J.-G.}\
  \bibnamefont {Zhu}}, \bibinfo {author} {\bibfnamefont {T.}~\bibnamefont
  {Kanbe}}, \ and\ \bibinfo {author} {\bibfnamefont {T.}~\bibnamefont
  {Shige}},\ }\href@noop {} {\bibfield  {journal} {\bibinfo  {journal} {J.
  Appl. Phys.}\ }\textbf {\bibinfo {volume} {113}},\ \bibinfo {pages} {17C107}
  (\bibinfo {year} {2013}{\natexlab{b}})}\BibitemShut {NoStop}%
\bibitem [{\citenamefont {Van~Driel}\ \emph {et~al.}(1999)\citenamefont
  {Van~Driel}, \citenamefont {Coehoorn}, \citenamefont {Strijkers},
  \citenamefont {Bruck},\ and\ \citenamefont {De~Boer}}]{van1999compositional}%
  \BibitemOpen
  \bibfield  {author} {\bibinfo {author} {\bibfnamefont {J.}~\bibnamefont
  {Van~Driel}}, \bibinfo {author} {\bibfnamefont {R.}~\bibnamefont {Coehoorn}},
  \bibinfo {author} {\bibfnamefont {G.}~\bibnamefont {Strijkers}}, \bibinfo
  {author} {\bibfnamefont {E.}~\bibnamefont {Bruck}}, \ and\ \bibinfo {author}
  {\bibfnamefont {F.}~\bibnamefont {De~Boer}},\ }\href@noop {} {\bibfield
  {journal} {\bibinfo  {journal} {J. Appl. Phys.}\ }\textbf {\bibinfo {volume}
  {85}},\ \bibinfo {pages} {1026} (\bibinfo {year} {1999})}\BibitemShut
  {NoStop}%
\bibitem [{\citenamefont {Cao}\ \emph {et~al.}(2008)\citenamefont {Cao},
  \citenamefont {Nam}, \citenamefont {Inoue}, \citenamefont {Ko}, \citenamefont
  {Phuoc},\ and\ \citenamefont {Suzuki}}]{cao2008magnetization}%
  \BibitemOpen
  \bibfield  {author} {\bibinfo {author} {\bibfnamefont {J.}~\bibnamefont
  {Cao}}, \bibinfo {author} {\bibfnamefont {N.~T.}\ \bibnamefont {Nam}},
  \bibinfo {author} {\bibfnamefont {S.}~\bibnamefont {Inoue}}, \bibinfo
  {author} {\bibfnamefont {H.~Y.~Y.}\ \bibnamefont {Ko}}, \bibinfo {author}
  {\bibfnamefont {N.~N.}\ \bibnamefont {Phuoc}}, \ and\ \bibinfo {author}
  {\bibfnamefont {T.}~\bibnamefont {Suzuki}},\ }\href@noop {} {\bibfield
  {journal} {\bibinfo  {journal} {J. Appl. Phys.}\ }\textbf {\bibinfo {volume}
  {103}},\ \bibinfo {pages} {07F501} (\bibinfo {year} {2008})}\BibitemShut
  {NoStop}%
\bibitem [{\citenamefont {Suzuki}\ \emph {et~al.}(2009)\citenamefont {Suzuki},
  \citenamefont {Koike}, \citenamefont {Itoh}, \citenamefont {Taniyama},\ and\
  \citenamefont {Sato}}]{suzuki2009stability}%
  \BibitemOpen
  \bibfield  {author} {\bibinfo {author} {\bibfnamefont {I.}~\bibnamefont
  {Suzuki}}, \bibinfo {author} {\bibfnamefont {T.}~\bibnamefont {Koike}},
  \bibinfo {author} {\bibfnamefont {M.}~\bibnamefont {Itoh}}, \bibinfo {author}
  {\bibfnamefont {T.}~\bibnamefont {Taniyama}}, \ and\ \bibinfo {author}
  {\bibfnamefont {T.}~\bibnamefont {Sato}},\ }\href@noop {} {\bibfield
  {journal} {\bibinfo  {journal} {J. Appl. Phys.}\ }\textbf {\bibinfo {volume}
  {105}},\ \bibinfo {pages} {07E501} (\bibinfo {year} {2009})}\BibitemShut
  {NoStop}%
\bibitem [{\citenamefont {Hillion}\ \emph {et~al.}(2013)\citenamefont
  {Hillion}, \citenamefont {Cavallin}, \citenamefont {Vlaic}, \citenamefont
  {Tamion}, \citenamefont {Tournus}, \citenamefont {Khadra}, \citenamefont
  {Dreiser}, \citenamefont {Piamonteze}, \citenamefont {Notling}, \citenamefont
  {Rusponi}, \citenamefont {Sato}, \citenamefont {Konno}, \citenamefont
  {Proux},\ and\ \citenamefont {Dupuis}}]{Hillion}%
  \BibitemOpen
  \bibfield  {author} {\bibinfo {author} {\bibfnamefont {A.}~\bibnamefont
  {Hillion}}, \bibinfo {author} {\bibfnamefont {A.}~\bibnamefont {Cavallin}},
  \bibinfo {author} {\bibfnamefont {S.}~\bibnamefont {Vlaic}}, \bibinfo
  {author} {\bibfnamefont {A.}~\bibnamefont {Tamion}}, \bibinfo {author}
  {\bibfnamefont {F.}~\bibnamefont {Tournus}}, \bibinfo {author} {\bibfnamefont
  {G.}~\bibnamefont {Khadra}}, \bibinfo {author} {\bibfnamefont
  {J.}~\bibnamefont {Dreiser}}, \bibinfo {author} {\bibfnamefont
  {C.}~\bibnamefont {Piamonteze}}, \bibinfo {author} {\bibfnamefont
  {F.}~\bibnamefont {Notling}}, \bibinfo {author} {\bibfnamefont
  {S.}~\bibnamefont {Rusponi}}, \bibinfo {author} {\bibfnamefont
  {K.}~\bibnamefont {Sato}}, \bibinfo {author} {\bibfnamefont {T.~J.}\
  \bibnamefont {Konno}}, \bibinfo {author} {\bibfnamefont {O.}~\bibnamefont
  {Proux}}, \ and\ \bibinfo {author} {\bibfnamefont {H.}~\bibnamefont {Dupuis},
  \bibfnamefont {V.~Brune}},\ }\href@noop {} {\bibfield  {journal} {\bibinfo
  {journal} {Phys. Rev. Lett.}\ }\textbf {\bibinfo {volume} {110}},\ \bibinfo
  {pages} {087207} (\bibinfo {year} {2013})}\BibitemShut {NoStop}%
\bibitem [{\citenamefont {Thiele}\ \emph {et~al.}(2004)\citenamefont {Thiele},
  \citenamefont {Maat}, \citenamefont {Robertson},\ and\ \citenamefont
  {Fullerton}}]{thiele2004magnetic}%
  \BibitemOpen
  \bibfield  {author} {\bibinfo {author} {\bibfnamefont {J.-U.}\ \bibnamefont
  {Thiele}}, \bibinfo {author} {\bibfnamefont {S.}~\bibnamefont {Maat}},
  \bibinfo {author} {\bibfnamefont {J.~L.}\ \bibnamefont {Robertson}}, \ and\
  \bibinfo {author} {\bibfnamefont {E.~E.}\ \bibnamefont {Fullerton}},\
  }\href@noop {} {\bibfield  {journal} {\bibinfo  {journal} {IEEE Trans.
  Magn.}\ }\textbf {\bibinfo {volume} {40}},\ \bibinfo {pages} {2537} (\bibinfo
  {year} {2004})}\BibitemShut {NoStop}%
\bibitem [{\citenamefont {Staunton}\ \emph {et~al.}(2014)\citenamefont
  {Staunton}, \citenamefont {Banerjee}, \citenamefont {dos Santos~Dias},
  \citenamefont {Deak},\ and\ \citenamefont {Szunyogh}}]{Staunton}%
  \BibitemOpen
  \bibfield  {author} {\bibinfo {author} {\bibfnamefont {J.~B.}\ \bibnamefont
  {Staunton}}, \bibinfo {author} {\bibfnamefont {R.}~\bibnamefont {Banerjee}},
  \bibinfo {author} {\bibfnamefont {M.}~\bibnamefont {dos Santos~Dias}},
  \bibinfo {author} {\bibfnamefont {A.}~\bibnamefont {Deak}}, \ and\ \bibinfo
  {author} {\bibfnamefont {L.}~\bibnamefont {Szunyogh}},\ }\href@noop {}
  {\bibfield  {journal} {\bibinfo  {journal} {Phys. Rev. B.}\ }\textbf
  {\bibinfo {volume} {89}},\ \bibinfo {pages} {054427} (\bibinfo {year}
  {2014})}\BibitemShut {NoStop}%
\bibitem [{\citenamefont {Cherifi}\ \emph {et~al.}(2014)\citenamefont
  {Cherifi}, \citenamefont {Ivanovskaya}, \citenamefont {Phillips},
  \citenamefont {Zobelli}, \citenamefont {Infante}, \citenamefont {Jacquet},
  \citenamefont {Garcia}, \citenamefont {Fusil}, \citenamefont {Briddon},
  \citenamefont {Guiblin}, \citenamefont {Mougin}, \citenamefont {Unal},
  \citenamefont {Kronast}, \citenamefont {Valencia}, \citenamefont {Dkhil},
  \citenamefont {Barthelemy},\ and\ \citenamefont {Bibes}}]{Cherify-14}%
  \BibitemOpen
  \bibfield  {author} {\bibinfo {author} {\bibfnamefont {R.}~\bibnamefont
  {Cherifi}}, \bibinfo {author} {\bibfnamefont {V.}~\bibnamefont
  {Ivanovskaya}}, \bibinfo {author} {\bibfnamefont {L.~C.}\ \bibnamefont
  {Phillips}}, \bibinfo {author} {\bibfnamefont {A.}~\bibnamefont {Zobelli}},
  \bibinfo {author} {\bibfnamefont {I.}~\bibnamefont {Infante}}, \bibinfo
  {author} {\bibfnamefont {E.}~\bibnamefont {Jacquet}}, \bibinfo {author}
  {\bibfnamefont {V.}~\bibnamefont {Garcia}}, \bibinfo {author} {\bibfnamefont
  {S.}~\bibnamefont {Fusil}}, \bibinfo {author} {\bibfnamefont
  {P.}~\bibnamefont {Briddon}}, \bibinfo {author} {\bibfnamefont
  {N.}~\bibnamefont {Guiblin}}, \bibinfo {author} {\bibfnamefont
  {A.}~\bibnamefont {Mougin}}, \bibinfo {author} {\bibfnamefont
  {A.}~\bibnamefont {Unal}}, \bibinfo {author} {\bibfnamefont {F.}~\bibnamefont
  {Kronast}}, \bibinfo {author} {\bibfnamefont {S.}~\bibnamefont {Valencia}},
  \bibinfo {author} {\bibfnamefont {B.}~\bibnamefont {Dkhil}}, \bibinfo
  {author} {\bibfnamefont {A.}~\bibnamefont {Barthelemy}}, \ and\ \bibinfo
  {author} {\bibfnamefont {M.}~\bibnamefont {Bibes}},\ }\href@noop {}
  {\bibfield  {journal} {\bibinfo  {journal} {Nature Mater.}\ }\textbf
  {\bibinfo {volume} {13}},\ \bibinfo {pages} {345} (\bibinfo {year}
  {2014})}\BibitemShut {NoStop}%
\bibitem [{\citenamefont {Junquera}\ and\ \citenamefont
  {Ghosez}(2003)}]{juncekranature}%
  \BibitemOpen
  \bibfield  {author} {\bibinfo {author} {\bibfnamefont {J.}~\bibnamefont
  {Junquera}}\ and\ \bibinfo {author} {\bibfnamefont {P.}~\bibnamefont
  {Ghosez}},\ }\href@noop {} {\bibfield  {journal} {\bibinfo  {journal}
  {Nature}\ }\textbf {\bibinfo {volume} {422}},\ \bibinfo {pages} {506}
  (\bibinfo {year} {2003})}\BibitemShut {NoStop}%
\bibitem [{\citenamefont {Garcia}\ and\ \citenamefont
  {Bibes}(2014)}]{garcia2014ferroelectric}%
  \BibitemOpen
  \bibfield  {author} {\bibinfo {author} {\bibfnamefont {V.}~\bibnamefont
  {Garcia}}\ and\ \bibinfo {author} {\bibfnamefont {M.}~\bibnamefont {Bibes}},\
  }\href@noop {} {\bibfield  {journal} {\bibinfo  {journal} {Nat. Commun.}\
  }\textbf {\bibinfo {volume} {5}},\ \bibinfo {pages} {4289} (\bibinfo {year}
  {2014})}\BibitemShut {NoStop}%
\bibitem [{\citenamefont {Rayson}\ and\ \citenamefont
  {Briddon}(2008)}]{Aimp-2}%
  \BibitemOpen
  \bibfield  {author} {\bibinfo {author} {\bibfnamefont {M.}~\bibnamefont
  {Rayson}}\ and\ \bibinfo {author} {\bibfnamefont {P.}~\bibnamefont
  {Briddon}},\ }\href@noop {} {\bibfield  {journal} {\bibinfo  {journal}
  {Comput. Phys. Commun.}\ }\textbf {\bibinfo {volume} {178}},\ \bibinfo
  {pages} {128} (\bibinfo {year} {2008})}\BibitemShut {NoStop}%
\bibitem [{\citenamefont {Jones}\ and\ \citenamefont {Briddon}(1998)}]{Aimpro}%
  \BibitemOpen
  \bibfield  {author} {\bibinfo {author} {\bibfnamefont {R.}~\bibnamefont
  {Jones}}\ and\ \bibinfo {author} {\bibfnamefont {P.}~\bibnamefont
  {Briddon}},\ }\href@noop {} {\bibfield  {journal} {\bibinfo  {journal}
  {Semicond. Semimetals}\ }\textbf {\bibinfo {volume} {51A}},\ \bibinfo {pages}
  {287} (\bibinfo {year} {1998})}\BibitemShut {NoStop}%
\bibitem [{\citenamefont {Hartwigsen}\ \emph {et~al.}(1998)\citenamefont
  {Hartwigsen}, \citenamefont {Goedecker},\ and\ \citenamefont
  {Hutter}}]{Hgh-98}%
  \BibitemOpen
  \bibfield  {author} {\bibinfo {author} {\bibfnamefont {C.}~\bibnamefont
  {Hartwigsen}}, \bibinfo {author} {\bibfnamefont {S.}~\bibnamefont
  {Goedecker}}, \ and\ \bibinfo {author} {\bibfnamefont {J.}~\bibnamefont
  {Hutter}},\ }\href@noop {} {\bibfield  {journal} {\bibinfo  {journal} {Phys.
  Rev. B}\ }\textbf {\bibinfo {volume} {58}},\ \bibinfo {pages} {3641}
  (\bibinfo {year} {1998})}\BibitemShut {NoStop}%
\bibitem [{\citenamefont {Monkhorst}\ and\ \citenamefont
  {Pack}(1976)}]{Monkhorst-Pack}%
  \BibitemOpen
  \bibfield  {author} {\bibinfo {author} {\bibfnamefont {H.~J.}\ \bibnamefont
  {Monkhorst}}\ and\ \bibinfo {author} {\bibfnamefont {J.~D.}\ \bibnamefont
  {Pack}},\ }\href {\doibase 10.1103/PhysRevB.13.5188} {\bibfield  {journal}
  {\bibinfo  {journal} {Phys. Rev. B}\ }\textbf {\bibinfo {volume} {13}},\
  \bibinfo {pages} {5188} (\bibinfo {year} {1976})}\BibitemShut {NoStop}%
\bibitem [{\citenamefont {Bocher}\ \emph {et~al.}(2011)\citenamefont {Bocher},
  \citenamefont {Gloter}, \citenamefont {Crassous}, \citenamefont {Garcia},
  \citenamefont {March}, \citenamefont {Zobelli}, \citenamefont {Valencia},
  \citenamefont {Enouz-Vedrenne}, \citenamefont {Moya}, \citenamefont {Marthur}
  \emph {et~al.}}]{bocher2011atomic}%
  \BibitemOpen
  \bibfield  {author} {\bibinfo {author} {\bibfnamefont {L.}~\bibnamefont
  {Bocher}}, \bibinfo {author} {\bibfnamefont {A.}~\bibnamefont {Gloter}},
  \bibinfo {author} {\bibfnamefont {A.}~\bibnamefont {Crassous}}, \bibinfo
  {author} {\bibfnamefont {V.}~\bibnamefont {Garcia}}, \bibinfo {author}
  {\bibfnamefont {K.}~\bibnamefont {March}}, \bibinfo {author} {\bibfnamefont
  {A.}~\bibnamefont {Zobelli}}, \bibinfo {author} {\bibfnamefont
  {S.}~\bibnamefont {Valencia}}, \bibinfo {author} {\bibfnamefont
  {S.}~\bibnamefont {Enouz-Vedrenne}}, \bibinfo {author} {\bibfnamefont
  {X.}~\bibnamefont {Moya}}, \bibinfo {author} {\bibfnamefont {N.~D.}\
  \bibnamefont {Marthur}},  \emph {et~al.},\ }\href@noop {} {\bibfield
  {journal} {\bibinfo  {journal} {Nano Lett.}\ }\textbf {\bibinfo {volume}
  {12}},\ \bibinfo {pages} {376} (\bibinfo {year} {2011})}\BibitemShut
  {NoStop}%
\bibitem [{Note1()}]{Note1}%
  \BibitemOpen
  \bibinfo {note} {We find a cell parameters of 2.92 {\r A} and 2.94 {\r A} for
  AFM FeRh and FM FeRh, respectively, in good agreement with experimental
  values\cite {Fallot, muldawer}}\BibitemShut {NoStop}%
\bibitem [{\citenamefont {Takahashi}\ and\ \citenamefont
  {Oshima}(1995)}]{Takahashi}%
  \BibitemOpen
  \bibfield  {author} {\bibinfo {author} {\bibfnamefont {M.}~\bibnamefont
  {Takahashi}}\ and\ \bibinfo {author} {\bibfnamefont {R.}~\bibnamefont
  {Oshima}},\ }\href@noop {} {\bibfield  {journal} {\bibinfo  {journal} {J.
  Phys. C8}\ ,\ \bibinfo {pages} {491}} (\bibinfo {year} {1995})}\BibitemShut
  {NoStop}%
\bibitem [{\citenamefont {Oshima}\ \emph {et~al.}(2003)\citenamefont {Oshima},
  \citenamefont {Hori}, \citenamefont {Kibata}, \citenamefont {Komatsu},\ and\
  \citenamefont {Kiritani}}]{Oshima}%
  \BibitemOpen
  \bibfield  {author} {\bibinfo {author} {\bibfnamefont {R.}~\bibnamefont
  {Oshima}}, \bibinfo {author} {\bibfnamefont {F.}~\bibnamefont {Hori}},
  \bibinfo {author} {\bibfnamefont {Y.}~\bibnamefont {Kibata}}, \bibinfo
  {author} {\bibfnamefont {M.}~\bibnamefont {Komatsu}}, \ and\ \bibinfo
  {author} {\bibfnamefont {M.}~\bibnamefont {Kiritani}},\ }\href@noop {}
  {\bibfield  {journal} {\bibinfo  {journal} {Mat. Sci. Eng.}\ }\textbf
  {\bibinfo {volume} {A350}},\ \bibinfo {pages} {139} (\bibinfo {year}
  {2003})}\BibitemShut {NoStop}%
\bibitem [{\citenamefont {Gruner}\ \emph {et~al.}(2003)\citenamefont {Gruner},
  \citenamefont {Hoffmann},\ and\ \citenamefont {Entel}}]{PhysRevB.67.064415}%
  \BibitemOpen
  \bibfield  {author} {\bibinfo {author} {\bibfnamefont {M.~E.}\ \bibnamefont
  {Gruner}}, \bibinfo {author} {\bibfnamefont {E.}~\bibnamefont {Hoffmann}}, \
  and\ \bibinfo {author} {\bibfnamefont {P.}~\bibnamefont {Entel}},\ }\href
  {\doibase 10.1103/PhysRevB.67.064415} {\bibfield  {journal} {\bibinfo
  {journal} {Phys. Rev. B}\ }\textbf {\bibinfo {volume} {67}},\ \bibinfo
  {pages} {064415} (\bibinfo {year} {2003})}\BibitemShut {NoStop}%
\bibitem [{\citenamefont {Sandratskii}\ and\ \citenamefont
  {Mavropoulos}(2011)}]{sandratskii2011magnetic}%
  \BibitemOpen
  \bibfield  {author} {\bibinfo {author} {\bibfnamefont {L.~M.}\ \bibnamefont
  {Sandratskii}}\ and\ \bibinfo {author} {\bibfnamefont {P.}~\bibnamefont
  {Mavropoulos}},\ }\href@noop {} {\bibfield  {journal} {\bibinfo  {journal}
  {Phys. Rev. B}\ }\textbf {\bibinfo {volume} {83}},\ \bibinfo {pages} {174408}
  (\bibinfo {year} {2011})}\BibitemShut {NoStop}%
\bibitem [{\citenamefont {Gu}\ and\ \citenamefont
  {Antropov}(2005)}]{gu2005dominance}%
  \BibitemOpen
  \bibfield  {author} {\bibinfo {author} {\bibfnamefont {R.}~\bibnamefont
  {Gu}}\ and\ \bibinfo {author} {\bibfnamefont {V.}~\bibnamefont {Antropov}},\
  }\href@noop {} {\bibfield  {journal} {\bibinfo  {journal} {Phys. Rev. B}\
  }\textbf {\bibinfo {volume} {72}},\ \bibinfo {pages} {12403} (\bibinfo {year}
  {2005})}\BibitemShut {NoStop}%
\bibitem [{\citenamefont {Miyajima}\ and\ \citenamefont
  {Yuasa}(1992)}]{Miyajima}%
  \BibitemOpen
  \bibfield  {author} {\bibinfo {author} {\bibfnamefont {H.}~\bibnamefont
  {Miyajima}}\ and\ \bibinfo {author} {\bibfnamefont {S.}~\bibnamefont
  {Yuasa}},\ }\href@noop {} {\bibfield  {journal} {\bibinfo  {journal} {J.
  Magn. Magn. Mater.}\ }\textbf {\bibinfo {volume} {104--107}},\ \bibinfo
  {pages} {2025} (\bibinfo {year} {1992})}\BibitemShut {NoStop}%
\bibitem [{\citenamefont {Duan}\ \emph {et~al.}(2006)\citenamefont {Duan},
  \citenamefont {Jaswal},\ and\ \citenamefont {Tsymbal}}]{duan2006predicted}%
  \BibitemOpen
  \bibfield  {author} {\bibinfo {author} {\bibfnamefont {C.-G.}\ \bibnamefont
  {Duan}}, \bibinfo {author} {\bibfnamefont {S.~S.}\ \bibnamefont {Jaswal}}, \
  and\ \bibinfo {author} {\bibfnamefont {E.~Y.}\ \bibnamefont {Tsymbal}},\
  }\href@noop {} {\bibfield  {journal} {\bibinfo  {journal} {Phys. Rev. Lett.}\
  }\textbf {\bibinfo {volume} {97}},\ \bibinfo {pages} {047201} (\bibinfo
  {year} {2006})}\BibitemShut {NoStop}%
\bibitem [{\citenamefont {Catalan}(2008)}]{catalan2008progress}%
  \BibitemOpen
  \bibfield  {author} {\bibinfo {author} {\bibfnamefont {G.}~\bibnamefont
  {Catalan}},\ }\href@noop {} {\bibfield  {journal} {\bibinfo  {journal} {Phase
  Transitions}\ }\textbf {\bibinfo {volume} {81}},\ \bibinfo {pages} {729}
  (\bibinfo {year} {2008})}\BibitemShut {NoStop}%
\bibitem [{\citenamefont {Medarde}(1997)}]{medarde1997structural}%
  \BibitemOpen
  \bibfield  {author} {\bibinfo {author} {\bibfnamefont {M.~L.}\ \bibnamefont
  {Medarde}},\ }\href@noop {} {\bibfield  {journal} {\bibinfo  {journal} {J.
  Phys.: Condens. Matter}\ }\textbf {\bibinfo {volume} {9}},\ \bibinfo {pages}
  {1679} (\bibinfo {year} {1997})}\BibitemShut {NoStop}%
\bibitem [{\citenamefont {Imada}\ \emph {et~al.}(1998)\citenamefont {Imada},
  \citenamefont {Fujimori},\ and\ \citenamefont {Tokura}}]{RevModPhys.70.1039}%
  \BibitemOpen
  \bibfield  {author} {\bibinfo {author} {\bibfnamefont {M.}~\bibnamefont
  {Imada}}, \bibinfo {author} {\bibfnamefont {A.}~\bibnamefont {Fujimori}}, \
  and\ \bibinfo {author} {\bibfnamefont {Y.}~\bibnamefont {Tokura}},\ }\href
  {\doibase 10.1103/RevModPhys.70.1039} {\bibfield  {journal} {\bibinfo
  {journal} {Rev. Mod. Phys.}\ }\textbf {\bibinfo {volume} {70}},\ \bibinfo
  {pages} {1039} (\bibinfo {year} {1998})}\BibitemShut {NoStop}%
\end{thebibliography}
\end{document}